\documentclass[12pt]{article}

\usepackage{graphicx}
\usepackage{latexsym}
\usepackage{fancyhdr}
\usepackage{subfigure}
\usepackage{mdwlist}

\newcommand{\Xbar}{{\overline X}}

\makeatletter

\def\noflash#1{\setbox0=\hbox{#1}\hbox to 1\wd0{\hfill}}

\newcommand{\comment}[1]{}

\newcommand{\nocomment}[1]{}

\newcommand{\Iitemize}{\begin{itemize}
	{\setlength{\itemsep}{-6pt}}
       }

\newcommand{\ls}[1]
   {\dimen0=\fontdimen6\the\font 
    \lineskip=#1\dimen0
    \advance\lineskip.5\fontdimen5\the\font
    \advance\lineskip-\dimen0
    \lineskiplimit=.9\lineskip
    \baselineskip=\lineskip
    \advance\baselineskip\dimen0
    \normallineskip\lineskip
    \normallineskiplimit\lineskiplimit
    \normalbaselineskip\baselineskip
    \ignorespaces
   }

\def\ifundefined#1{\expandafter\ifx\csname#1\endcsname\relax}
\ifundefined{newblock}{
 }\fi

\newcommand{\eqref}[1]{Equation~\ref{#1}}

\begin{document}

\title{Deterministic Consensus Algorithm\\ with Linear Per-Bit Complexity  \footnote{\normalsize This research is supported
in part by Army Research Office grant
W-911-NF-0710287. Any opinions, findings, and conclusions or recommendations expressed here are those of the authors and do not
necessarily reflect the views of the funding agencies or the U.S. government.}}

\date{\today}
\author{Guanfeng Liang and Nitin Vaidya\\ \normalsize Department of Electrical and Computer Engineering, and\\ \normalsize Coordinated Science Laboratory\\ \normalsize University of Illinois at Urbana-Champaign\\ \normalsize gliang2@illinois.edu, nhv@illinois.edu\\~\\Technical Report}

%


\maketitle


\thispagestyle{empty}

\newpage

\setcounter{page}{1}

\section{Introduction}
In this report, building on the deterministic multi-valued one-to-many Byzantine agreement ({\bf broadcast}) algorithm in our recent technical report
 \cite{techreport_BA_complexity}, we introduce a deterministic multi-valued all-to-all Byzantine agreement algorithm ({\bf consensus}), with linear complexity per bit agreed upon. The discussion in this note is not self-contained, and relies heavily on the material in \cite{techreport_BA_complexity} -- please refer to
\cite{techreport_BA_complexity} for the necessary background.

Consider a synchronous fully connected network with $n$ nodes, namely $0,1,\dots, n-1$. At most $t<n/3$ nodes can be faulty. Every node $i$ is given an initial value of $L$ bits. The goal of a {\em consensus} algorithm is to allow each node to decide (or agree) on a value consisting of $L$ bits,
 while satisfying the following three requirements:

\begin{itemize}
\item Every fault-free node eventually decides on a value (termination);
\item The decided values of all fault-free nodes are equal (consistency);
\item If every fault-free node holds the same initial value $v$, the decided value equals to $v$ (validity).
\end{itemize}

Our algorithm achieves consensus on a long value of $L$ bits deterministically. Similar to the one-to-many algorithm in \cite{techreport_BA_complexity}, the proposed all-to-all Byzantine agreement (or {\em consensus})
 algorithm  progresses in generations. In each generation, $D$ bits are being agreed upon, with the total number of generations being $L/D$.
For convenience, we assume $L$ to be an integral multiple of $D$.

\section{Consensus Algorithm}\label{sec:algorithm}
In the proposed {\em consensus}\, algorithm, we use the same technique of ``diagnosis graph'' to narrow down the locations of faulty nodes as in \cite{techreport_BA_complexity}. If a node $y$ is accused by at least $t+1$ other nodes, $y$ must be faulty. Then it is isolated,
and does not perform the algorithm below. When a new node is isolated,
essentially $n$ decreases by 1, and $t$ also decreases by 1. For the
reduced network, the condition that $n>3t$ will continue to hold, if
it held previously. In the
following, we consider the reduced network with the reduced values of
$n,t$, and assume that no node in the reduced network is accused by $>t$ nodes.
When we say ``network'', we mean the reduced network below.

The following steps are performed for the $D$ bits of information of the current generation.
Let the $D$ bits at node $i$ be denoted as $v_i$.
\begin{enumerate}

\item This step is performed by each node $i$:
We use a $(n,n-2t)$ distance-$(2t+1)$ code, wherein each codeword
consists of $n$ symbols, each symbol of size $D/(n-2t)$ bits.
Such a code exists, provided that the symbol size is large enough.
Let us denote this code as $C_{2t}$.
With a symbol size of $D/(n-2t)$ bits, the $D$-bit value at node $i$
can be viewed as $(n-2t)$ symbols. Encode $v_i$ into a
codeword $s_i$ from the code $C_{2t}$.
The $j$-th symbol in the codeword is denoted as $s_{ij}$.
Send $s_{ii}$ to all other nodes that it trusts.
Thus, node $i$ sends $i$-th symbol of its codeword
to all nodes {\em that it trusts}.

{\bf Note for future reference:}
Since code $C_{2t}$ has distance $2t+1$,
any punctured $(n-z, n-2t)$ code obtained from $C_{2t}$
has distance $2t+1-z$, where $z\leq 2t$. Let $C_t$ denote the
punctured $(n-t, n-2t)$ code of distance $t+1$ obtained by removing the
last $t$ symbols of the original $(n,n-2t)$ code above.
By ``last'' $t$ symbols, we refer to symbols with index $n-t-1$ through $n-1$.

\item This step is performed by each node $i$:
Denote by $r_{ij}$ the symbol received from node $j$ in step 1.
If $i$ trusts $j$ and $r_{ij}=s_{ij}$, then set $M_{ij}=TRUE$; else $M_{ij}=FALSE$.
$M_i$ is a ``match'' vector, and records whether $i$'s information matches
with the symbols sent by the other nodes.

\item Each node $i$ uses traditional Byzantine agreement (one-to-many) algorithm
to broadcast $M_i$ to all the nodes. 

\item Now each node $i$ has received $M_j$ from each node $j$.
Due to the use of BA in the previous step, all fault-free nodes
receive identical $M$ vectors.
Each node $i$ attempts to find a set $X$ containing exactly $(n-t)$ nodes that are
``collectively consistent''. That is, for every pair
of nodes $j,k\in X$, $M_{jk}=M_{kj}=TRUE$.

There are two cases:
\begin{itemize}
\item No such subset $X$ exists: Note that if all fault-free nodes
(at least $n-t$ of them exist) have identical initial value, then a set
$X$ must exist. (Fault-free nodes always trust each other.)
Thus, if no such $X$ exists, that implies that the
fault-free nodes do not have identical value. Thus, the fault-free
nodes can agree on a default value, and terminate the algorithm.

\item At least one such subset $X$ exists:
In this case, all fault-free nodes identify one such set $X$ using
a deterministic algorithm (thus, all nodes should identify the same $X$).
Since all fault-free nodes can compute $X$ identically,
without loss of generality, suppose that $X$ contains nodes
0 through $(n-t-1)$. Thus, the nodes {\em not}\, in $X$ are $n-t$ through $n-1$.
(In other words, the nodes are renumbered after $X$ is computed.)
Thus, 
\[ X = \{ 0, 1, \cdots, n-t-1\} \]
and define
\[ \Xbar = \{n-t, \cdots, n-1\} \]
Let the $(n-t)$-symbol received vector at node $i$ consisting of the
symbols received from the $(n-t)$ nodes in $X$ be called $R_i$.\footnote{In vector $R_i$,
the symbols are arranged in increasing order of the identifiers of the nodes that sent them.}

Since $X$ contains $n-t$ nodes and there are at most $t$ faulty nodes, at least $n-2t\ge 2$ of these nodes
must be fault-free.
Consider two fault-free nodes $j$ and $k$ in $X$.
By definition of $R_j$ and $R_k$, nodes $j$ and $k$
find these vectors ``consistent'' with their own values
$v_j$ and $v_k$, respectively. In other words,
$R_j$ and $R_k$ are codewords in $C_t$.

There are at least $n-2t$  fault-free nodes in $X$, which must
have sent the same symbols to nodes $j$ and $k$ in step 1.
Thus, the $(n-t)$-symbol vectors $R_j$ and $R_k$ must be identical
in at least $(n-2t)$ positions, and differ in at most $t$ positions.

Given that (i) $C_t$ is a distance $t+1$ code, (ii) $R_j$
and $R_k$ are both codewords in $C_t$, and (iii) $R_j$
and $R_k$ differ in at most $t$ positions, it follows that
$R_j$ and $R_k$ must be identical. This, in turn, implies that
$v_j$ and $v_k$ must be identical as well.
This proves the following claim:

~

{\bf Claim 1:} All fault-free nodes in $X$ have identical $D$-bit
values. In other words, for all fault-free nodes $ j,k \in X$, 
$v_j=v_k$.

\end{itemize}

\item \label{step_y} Now consider a node $y \in \Xbar$.
Identify any node $z_y$ in $X$ such that $z_y$ and $y$ trust each other.
If no such $z_y$ exists, that implies that $y$ is accused by all
$n-t>t$ nodes in $X$, and therefore, $y$ must be already identified
as faulty, and must have
been isolated previously.
Thus, $z_y$ exists.

For each $y\in \Xbar$,
node $z_y$ transmits $t$ symbols $s_{z_y(n-t)}$ through $s_{z_y(n-1)}$ to node
$y$.\footnote{For complexity analysis presented later, note that there are $t$ nodes in $\Xbar$,
each of which is sent $t$ symbols each consisting of $\frac{D}{n-2t}$ bits.}
Each fault-free node $y\in \Xbar$ 
forms a vector using the $(n-t)$ symbols $r_{j0}\cdots r_{j(n-t-1)}$
received in step 1, and the above $t$ symbols received from node $z_y$.
Suppose that the $n$-symbol vector thus formed at fault-free node $y$ is
denoted $F_y$.

\paragraph{Failure detection rule:}
If $F_y$ is {\em not}\, a valid codeword from the $(n,n-2t)$ code $C_{2t}$, then
node $y$ detects a failure. This failure observation is distributed
to other nodes in the next step. (Justification for this failure detection
mechanism is presented below.)

\item All nodes in $\Xbar$ broadcast (using a traditional BA algorithm)
a single bit notification announcing whether they detected a failure
in the above step.

\paragraph{Decision rule:}
If no failure detection is announced by anyone, then each fault-free node $i$ in $X$
decides (agrees) on its own value $v_i$, and each fault-free node $j\in \Xbar$ decides
on the value corresponding to the codeword $F_j$.

If a failure is detected by anyone, then the failure is narrowed down using
a ``full broadcast'' procedure described in \cite{techreport_BA_complexity}, and agreement on the $D$ bits is also achieved
as a part of this full broadcast. The diagnosis graph is updated, and we return
to step 1 for next set of $D$ bits. 

\end{enumerate}

\paragraph{Justification for the failure detection and decision rules:}
Consider any fault-free node $i\in X$ and any fault-free node $y\in \Xbar$.
Now let us compare $F_y$ with $s_i$.
\begin{itemize}
\item Consider the first $n-t$ symbols of these vectors (elements with
index 0 through $n-t-1$). Observe that, for fault-free node $i\in X$,
$r_{ij}=s_{ij}$, for $0\leq j\leq n-t-1$, by definition of set $X$.
Since at least $n-2t$ of the symbols with index $<n-t$ come from
fault-free nodes, $F_y$ can differ from $r_i$ (and $s_i$) in at most
$t$ positions with index $<n-t$.
\item Consider the last $t$ symbols of vectors $F_y$ and $s_i$.
Since $z_y$ may be faulty and could have sent arbitrary $t$ symbols to 
to $y$ in step~\ref{step_y}, vectors $F_y$ and $s_i$ may differ
in all of these $t$ positions.
\end{itemize}
Thus, $r_i$ and $F_j$ may differ in at most $2t$ positions. 
Now let us make two observations:
\begin{itemize}
\item
Observation 1:
By definition of $s_i$, $s_i$ is a valid codeword from the
distance-$(2t+1)$ code $C_{2t}$. 
Since $F_y$ differs from valid codeword $s_i$ only in $2t$ places,
it follows that: either (i) $s_i=F_y$
(and both are valid codewords), or (ii) $F_y$ is not a valid codeword.

\item
{Observation 2:}
To derive this observation,
consider the case where all the nodes in $X$ are fault-free.
Clearly, in this case, all these nodes must have same value (from
claim 1 above). Then $s_i$ is identical for all $i\in X$,
and thus the (fault-free) nodes in X send symbols consistent with
the common value in step 1 to the nodes in $\Xbar$.
$F_y$ ($y\in\Xbar$) consists entirely of symbols sent to it
by nodes in $X$. Thus, clearly, $F_y$ will be equal to $s_i$ for all
$i\in X$ when all nodes in $X$ are fault-free.
It then follows that $F_y$ is a codeword from the $(n,n-2t)$ code
$C_{2t}$
when all nodes in $X$ are fault-free.
\end{itemize}
The above two observations imply that: (i) if $F_y$ is not
a codeword then all the nodes in $X$ cannot be fault-free (that is,
at least one of these nodes must have behaved incorrectly), and
(ii) if $F_y$ is a codeword, then the value corresponding to $F_y$
matches with the values at all the fault-free nodes in $X$.  

Thus, when the failure detection rule above detects a failure,
a failure must have indeed occurred. Also, while Claim 1
shows that the fault-free nodes in $X$ will agree with each other
using the above decision rule, observation 1 implies that
fault-free nodes in $\Xbar$ will also agree with them.

\section{Complexity Analysis}\label{sec:complexity}
We have finished describing the proposed consensus algorithm above. Now let us study the communication complexity of this algorithm.

\begin{itemize}
\item In step 1, every node sends at most $n-1$ symbols of $D/(n-2t)$ bits. So at most 
\begin{equation}
\frac{n(n-1)}{n-2t}D~bits
\end{equation}
are transmitted. Notice that this value decreases when both $n$ and $t$ are reduces by the same amount. As a result, no more than $\frac{n(n-1)}{n-2t}D$ bits will be transmitted in step 1 when some nodes are isolated.

\item In step 3, every node broadcasts a ``match'' vector of  $n-1$ bits, using a traditional Byzantine agreement (one-to-many) algorithm. Let us denote $B$ as the bit-complexity to broadcast 1 bit. So in step 3, at most
\begin{equation}
n(n-1)B~bits
\end{equation}
are transmitted.

\item If no $X$ is found in step 4, the algorithm terminates and nothing is transmitted any more. So we only consider the case when $X$ exists. As we have seen before, in step \ref{step_y}, every node in $\Xbar$ receives $t$ symbols of $D/(n-2t)$ bits, which results in $\frac{t^2}{n-2t}D$ bits being transmitted. 
Additionally, in step 6, every node in $\Xbar$ broadcasts a 1-bit notification, which requires $tB$ bits being transmitted. So if no failure is detected, at most
\begin{equation}
\frac{t^2}{n-2t}D + tB~bits
\end{equation}
are transmitted in steps 5 and 6. Again, this value decreases when both $n$ and $t$ are reduced by the same amount. So if some nodes are isolated, fewer bits will be transmitted.

\item If a failure is detected in step 6, every node broadcasts all symbols it has sent and has received through steps 1 to \ref{step_y}. In step 1, $n(n-1)$ symbols are transmitted. In step \ref{step_y}, $t^2$ symbols are transmitted. So $2(n(n-1)+t^2)$ symbols are being broadcast after a failure is detected, which results in 
\begin{equation}
\frac{2(n(n-1)+t^2)}{n-2t}DB~bits
\end{equation}
being transmitted.  Again, this value decreases when both $n$ and $t$ are reduced by the same amount. So if some nodes are isolated, fewer bits will be transmitted.
\end{itemize}

Now we can compute an upper bound of the complexity of the proposed algorithm. Notice that $D$ bits are being agreed on in every generation, so there are $L/D$ generations in total\footnote{To simplify the presentation, we assume that $L$ is an integer multiple of $D$ here. For other values of $L$, the analysis and results are still valid by applying the ceiling function $\lceil\cdot\rceil$ to the number of generations.}. Thus, excluding the broadcast after failures are detected, no more than
\begin{eqnarray}
&&\left(\frac{n(n-1)}{n-2t}D + n(n-1)B + \frac{t^2}{n-2t}D + tB\right)\frac{L}{D}\\
&=& \frac{n(n-1)+t^2}{n-2t}L + \frac{(n(n-1)+t)BL}{D}~bits
\end{eqnarray}
are transmitted. In addition, similar to our one-to-many algorithm, all faulty nodes will be identified after failures are detected in at most $(t+1)t$ generations. So the ``full broadcast'' will be performed at most $(t+1)t$ times throughout the whole execution of the algorithm. So the total number of bits transmitted in the ``full broadcast'' in all generations is at most 
\begin{equation}
\frac{2(n(n-1)+t^2)(t+1)t}{n-2t}DB~bits
\end{equation}
An upper bound on the communication complexity of the proposed algorithm, denoted as $C(L)$ is then computed as
\begin{eqnarray}
C(L) \le \frac{n(n-1)+t^2}{n-2t}L + \frac{(n(n-1)+t)BL}{D} + \frac{2(n(n-1)+t^2)(t+1)t}{n-2t}DB
\end{eqnarray}
For a large enough value of $L$, with a suitable choice of 
\begin{eqnarray}
D = \sqrt{\frac{(n(n-1)+t)(n-2t)L}{2(n(n-1)+t^2)(t+1)t}},
\end{eqnarray}
we have
\begin{eqnarray}
C(L) &\le& \frac{n(n-1)+t^2}{n-2t}L + 2BL^{0.5}\sqrt{\frac{2(n(n-1)+t)(n(n-1)+t^2)(t+1)t}{n-2t}}\\
&=& \frac{n(n-1)+t^2}{n-2t}L + BL^{0.5}\Theta(n^{2.5})
\end{eqnarray}

Notice that deterministic broadcast algorithms of complexity $\Theta(n^2)$ are known \cite{bit_optimal_89}, so we assume $B=\Theta(n^2)$. Then the complexity of our algorithm for all $t<n/3$ is upper bounded by
\begin{equation}
C(L) \le \frac{n(n-1)+t^2}{n-2t}L + L^{0.5}\Theta(n^{4.5}) \le \frac{10}{3} nL + L^{0.5}\Theta(n^{4.5}).
\end{equation}

For a given network with size $n$, the per-bit communication complexity of our algorithm is upper bounded by
\begin{eqnarray}
\alpha(L) &=& \frac{C(L)}{L} \\
&\le& \frac{n(n-1)+t^2}{n-2t} + L^{-0.5}\Theta(n^{4.5}) \\
&\rightarrow& \frac{n(n-1)+t^2}{n-2t} = \Theta(n),~as~L\rightarrow\infty.
\end{eqnarray}

\bibliographystyle{abbrv}
\bibliography{PaperList}

\begin{thebibliography}{1}

\bibitem{bit_optimal_89}
P.~Berman, J.~A. Garay, and K.~J. Perry.
\newblock Bit optimal distributed consensus.
\newblock {\em Computer science: research and applications}, 1992.

\bibitem{techreport_BA_complexity}
G.~Liang and N.~Vaidya.
\newblock Complexity of multi-valued byzantine agreement.
\newblock {\em Technical Report, CSL, UIUC}, June 2010.

\end{thebibliography}

\end{document}